\documentclass{article}
\usepackage{cite}
\usepackage{graphicx}
\usepackage{dcolumn}

\begin{document}

\date{}
\title{On the two-dimensional hydrogen atom in a circular box in the presence of an
electric field}
\author{Paolo Amore\thanks{%
e--mail: paolo@ucol.mx} \\
%EndAName
Facultad de Ciencias, CUICBAS, Universidad de Colima,\\
Bernal D\'{\i}az del Castillo 340, Colima, Colima,Mexico \\
and \\
Francisco M. Fern\'andez\thanks{%
e--mail: fernande@quimica.unlp.edu.ar} \\
INIFTA, DQT,\\
Blvd. 113 y 64 (S/N), Sucursal 4, Casilla de Correo 16,\\
1900 La Plata, Argentina}
\maketitle

\begin{abstract}
We revisit the quantum-mechanical two-dimensional hydrogen atom with an
electric field confined to a circular box of impenetrable wall. In order to
obtain the energy spectrum we resort to the Rayleigh-Ritz method with a
polynomial basis sets. We discuss the limits of large and small box radius
and the symmetry of the solutions of the Schr\"{o}dinger equation. An
interesting feature of the model is the appearance of accidental degeneracy
and the splitting of degenerate energy levels due to the presence of the
electric field.
\end{abstract}

\section{Introduction}

\label{sec:intro}

Quantum mechanical models of particles confined within boxes of
different shapes have received considerable attention for many
years\cite{SBC09a,SBC09b,S14}. In such reviews one can find all
kind of atomic and molecular systems enclosed inside surfaces that
are impenetrable or penetrable. One of the simple models is that
of a hydrogen atom within an impenetrable box with an electric
field. In the case of paraboloidal surfaces the problem is
particularly simple because the problem is separable in
paraboloidal coordinates\cite{AFC82}. The same kind of coordinates
were proved useful for other kinds of surfaces\cite{FRT84}.

Here, we are interested in the case of a two-dimensional hydrogen atom
within a circular box with impenetrable wall under the effect of an electric
field. Longo et al\cite{LGL23} considered the nucleus clamped at origin as
well as at other points with the argument that the nucleus should be located
at the minimum of the potential-energy surface (PES). However, if the
nucleus moves one should consider its kinetic energy and the calculation
becomes somewhat more complicated\cite{F10} (see also chapter 3 in\cite{S14}
for a more rigorous approach). Longo et al resorted to the Born-Oppenheimer
approximation to justify the location of the nucleus at such points.
However, this approximation requires the solution of the Schr\"{o}dinger
equation for the nucleus moving on such PES\cite{BH54} which Longo et al did
not carry out. In other words, the location of the nucleus at the minimum of
the PES does not capture the whole physics of the problem because the
nucleus is still considered immovable.

In this paper we consider the nucleus clamped at origin because it
facilitates the calculation. It is our purpose to derive some analytical
results that have not been considered by Longo et al and also provide higly
accurate energy eigenvalues by means of the Rayleigh-Ritz method (RRM)\cite
{P68,SO96}. The main advantage of the RRM is that the approximate
eigenvalues approach the exact ones from above\cite{M33} (see also \cite
{F25a} and the references therein).

In section~\ref{sec:model} we discuss some analytical features of the model.
In section~\ref{sec:RR} we describe the basis set for the RRM calculations
discussed in sections \ref{sec:f=0} and \ref{sec:f>0}). Finally, in section~%
\ref{sec:conclusions} we summarize the main results and draw conclusions.

\section{The model}

\label{sec:model}

In this section we present the model and discuss some of the properties of
the time-independent Schr\"{o}dinger equation. More precisely, we are
interested in the eigenvalue equation $H\psi =E\psi $ for the Hamiltonian
operator
\begin{equation}
H=-\frac{\hbar ^{2}}{2m_{e}}\nabla ^{2}-\frac{K}{r}-fex,  \label{eq:H}
\end{equation}
where $m_{e}$ is the electron mass, $K>0$ is the strength of the Coulomb
potential, $f$ is the magnitude of the electric field, $-e$ is the electron
charge and $r=\sqrt{x^{2}+y^{2}}$. For simplicity, we assume that the
nucleus is clamped at the origin. The solutions $\psi (r,\phi )$ in polar
coordinates $x=r\cos \phi $, $y=r\sin \phi $, $0\leq \phi <2\pi $, satisfy
the boundary condition $\psi \left( r_{0},\phi \right) =0$ that is
determined by the impenetrable wall of a circular box of radius $r_{0}$.
Therefore, $0<r\leq r_{0}$.

In order to discuss some analytical properties of the solutions to the
eigenvalue equation it is convenient to carry out the transformation $%
(x,y)\rightarrow (L\tilde{x},L\tilde{y})$, $r\rightarrow L\tilde{r}$, $%
\nabla ^{2}\rightarrow L^{-2}\tilde{\nabla}^{2}$, where $L>0$ is a real
constant with units of length. In this way we derive a useful dimensionless
eigenvalue equation $\tilde{H}\tilde{\psi}=\tilde{E}\tilde{\psi}$, where\cite
{F20}
\begin{equation}
\tilde{H}=\frac{m_{e}L^{2}}{\hbar ^{2}}H=-\frac{1}{2}\tilde{\nabla}^{2}-%
\frac{m_{e}LK}{\hbar ^{2}\tilde{r}}-\frac{m_{e}feL^{3}}{\hbar ^{2}}\tilde{x}.
\label{eq:H_dim_L}
\end{equation}
The boundary condition becomes $\tilde{\psi}\left( \tilde{r}_{0},\phi
\right) =0$, where $\tilde{r}_{0}=r_{0}/L$.

The scaling properties of the Hamiltonian operator prove useful for the
derivation of many properties of quantum-mechanical systems as discussed
elsewhere\cite{F20}. In what follows we add another example.

If we choose $L=r_{0}$ then
\begin{equation}
\tilde{H}=-\frac{1}{2}\tilde{\nabla}^{2}-\frac{m_{e}r_{0}K}{\hbar ^{2}\tilde{%
r}}-\frac{m_{e}fer_{0}^{3}}{\hbar ^{2}}\tilde{x},  \label{eq:H_dim_L=r0}
\end{equation}
and $\tilde{\psi}(1,\phi )=0$. Therefore,
\begin{equation}
\lim\limits_{r_{0}\rightarrow 0}\tilde{H}=-\frac{1}{2}\tilde{\nabla}^{2},
\label{eq:H_dim_r0->0}
\end{equation}
and the problem reduces to a particle within a circular box of radius $%
\tilde{r}_{0}=1$.

For some of our present calculations we also resort to the unit of length $%
L=\hbar ^{2}/(m_{e}K)$ that leads to
\begin{equation}
\tilde{H}=-\frac{1}{2}\tilde{\nabla}^{2}-\frac{1}{\tilde{r}}-\lambda \tilde{x%
},\;\lambda =\frac{fe\hbar ^{4}}{m_{e}^{2}K^{3}},  \label{eq:H_dim_LK}
\end{equation}
and omit the tilde over the dimensionless variables from now on. One just
keeps in mind that the unit of energy is $\hbar
^{2}/(m_{e}L^{2})=m_{e}K^{2}/\hbar ^{2}$. More precisely, in what follows we
restrict ourselves to the dimensionless Hamiltonian operator
\begin{equation}
H=-\frac{1}{2}\nabla ^{2}-\frac{1}{r}-\lambda r\cos \phi .
\label{eq:H_dim_LK2}
\end{equation}

Before proceeding with our discussion of the model (\ref{eq:H_dim_LK2}) we
want to point out that expressions like ``we use units such that $\hbar =m=1$%
'' are meaningless if we do not clearly indicate the units of length and
energy actually used. Note that the two dimensionless Hamiltonians $\tilde{H}
$ shown in equations (\ref{eq:H_dim_L=r0}) and (\ref{eq:H_dim_LK}) formally
correspond to choosing $\hbar =m=1$ though the units of length and energy in
one case are different from those in the other.

The Hamiltonian operator (\ref{eq:H_dim_LK2}) is invariant under the unitary
transformation $U_{1}^{\dagger }\phi U_{1}=-\phi $. Therefore, the solutions
to the Schr\"{o}dinger equation are either even ($\psi ^{(e)}(r,-\phi )=\psi
^{(e)}(r,\phi )$) or odd ($\psi ^{(o)}(r,-\phi )=-\psi ^{(o)}(r,\phi )$) and
we can treat them separately. Each set of solutions is nondegenerate for all
values of $r_{0}>0$ and $-\infty <\lambda <\infty $, except for accidental
degeneracies.

Another useful unitary transformation is given by $U_{2}^{\dagger }\phi
U_{2}=\phi +\pi $ that leads to $U_{2}^{\dagger }H(\lambda )U_{2}=H(-\lambda
)$. It follows from
\begin{equation}
U_{2}^{\dagger }H(\lambda )\psi =U_{2}^{\dagger }H(\lambda
)U_{2}U_{2}^{\dagger }\psi =H(-\lambda )U_{2}^{\dagger }\psi
=EU_{2}^{\dagger }\psi ,
\end{equation}
that the eigenvalues are even functions of $\lambda $: $E(-\lambda
)=E(\lambda )$.

For $\lambda =0$ the energy eigenvalues can be labelled by the radial
quantum number $n=0,1,\ldots $ and the magnetic one $m=0,\pm 1,\pm 2,\ldots $%
. Since the energy eigenvalues do not depend on the sign of $m$ we may write
them as $E_{n\nu }$, where $\nu =|m|$. For finite values of $r_{0}$ the
eigenvalues are at least two-fold degenerate when $\nu >0$. On the other
hand, the degeneracy of the free atom eigenvalues is $(n+\nu +1)$ because
\begin{equation}
\lim\limits_{r_{0}\rightarrow \infty }E_{n\nu }=-\frac{2}{(2n+2\nu +1)^{2}}.
\label{eq:E_HA}
\end{equation}

Although $\nu $ is no longer a true quantum number when $\lambda \neq 0$ we
sometimes keep the same notation for the eigenvalues in order to facilitate
the discussion of the spectrum. This convention is useful because we can
predict, for example, that $E_{n0}\left( r_{0},0\right) \rightarrow
E_{n0}^{(e)}\left( r_{0},\lambda \right) $ and $E_{n\nu }\left(
r_{0},0\right) \rightarrow \left\{ E_{n\nu }^{(e)}\left( r_{0},\lambda
\right) ,E_{n\nu }^{(o)}\left( r_{0},\lambda \right) \right\} $ as shown in
the calculations below.

It is worth comparing present analysis with that one put forward recently
for the harmonic oscillator in a two dimensional circular box\cite{FGAF24}.
One of the main differences is the limit $r_{0}\rightarrow \infty $ and
other ones will be discussed below.

\section{Rayleigh-Ritz method}

\label{sec:RR}

For the even and odd eigenfunctions of the Hamiltonian operator (\ref
{eq:H_dim_LK2}) we propose the simple non-orthogonal basis sets
\begin{eqnarray}
f_{i,j}^{(e)}(r,\phi ) &=&r^{i}(r_{0}-r)\cos (j\phi ),\;i=0,1,\ldots
,\;j=0,1,\ldots ,i,  \nonumber \\
f_{i,j}^{(o)}(r,\phi ) &=&r^{i}(r_{0}-r)\sin (j\phi ),\;i=1,2,\ldots
,\;j=1,2,\ldots ,i,  \label{eq:f^eo}
\end{eqnarray}
respectively. The calculation of the matrix elements $S_{iji^{\prime
}j^{\prime }}=\left\langle f_{ij}\right| \left. f_{i^{\prime }j^{\prime
}}\right\rangle $ and $H_{iji^{\prime }j^{\prime }}=\left\langle
f_{ij}\right| H\left| f_{i^{\prime }j^{\prime }}\right\rangle $ is
straightforward and we can easily apply the RRM\cite{P68,SO96,M33,F24} after
converting $S_{iji^{\prime }j^{\prime }}$ and $H_{iji^{\prime }j^{\prime }}$
into two-dimensional arrays. The main disadvantage of these basis sets is
that they are not expected to be suitable for too large values of $r_{0}$
because they do not exhibit the correct exponential behaviour when $%
r_{0}\rightarrow \infty $. The larger the value of $r_{0}$ the greater the
dimension of the basis set for a given accuracy. However, they are suitable
for present discussion of the model.

We choose $j=0,1,\ldots ,N$, $i=j,j+1,\ldots ,N$ for even states and $%
j=1,2,\ldots ,N$, $i=j,j+1,\ldots ,N$ for odd ones.\ Consequently, the
number of basis functions is $(N+1)(N+2)/2$ in the former case and $N(N+1)/2$
in the latter. This basis set is identical to one of the basis sets used for
the study of the harmonic oscillator\cite{FGAF24}.

According to equation (\ref{eq:H_dim_r0->0}) we have
\begin{equation}
\lim\limits_{r_{0}\rightarrow 0}r_{0}^{2}E_{n\nu }\left( r_{0},\lambda
\right) =E_{n\nu }^{PB},  \label{eq:E_r0->0}
\end{equation}
where $E_{n\nu }^{PB}$ are the eigenvalues of the particle in a circular box
with radius $r_{0}=1$. Table~\ref{tab:r0^2E} shows $r_{0}^{2}E_{n\nu }\left(
r_{0},1\right) $ for two small values of $r_{0}$. Such numerical results
just confirm equation (\ref{eq:E_r0->0}). Although the eigenvalues $E_{n\nu
}\left( r_{0},\lambda \right) $ obtained in this paper are different from
those for the Harmonic oscillator\cite{FGAF24} the limit (\ref{eq:E_r0->0})
is identical.

\section{On the case $f=0$}

\label{sec:f=0}

In this section we briefly address the case $f=0$ by means of the
dimensionless Hamiltonian operator (\ref{eq:H_dim_L=r0}) that we can rewrite
as
\begin{equation}
H=-\frac{1}{2}\nabla ^{2}-\frac{\beta }{r},\;\beta =\frac{m_{e}r_{0}K}{\hbar
^{2}},  \label{eq:H_dim_L=r0_f=0}
\end{equation}
where we have omitted the tilde over the dimensional quantities.

The Schr\"{o}dinger equation for this model is separable in polar
coordinates $\psi (r,\phi )=R(r)e^{im\phi }$ and the boundary condition is $%
R(1)=0$. Since it only remains to solve the eigenvalue equation for the
radial part we resort to the simpler non-orthogonal basis set
\begin{equation}
f_{i\nu }(r)=r^{i+\nu }(1-r),\;i=0,1,\ldots .  \label{eq:f_(inu)_f=0}
\end{equation}

\begin{figure}[tbp]
\begin{center}
\includegraphics[width=9cm]{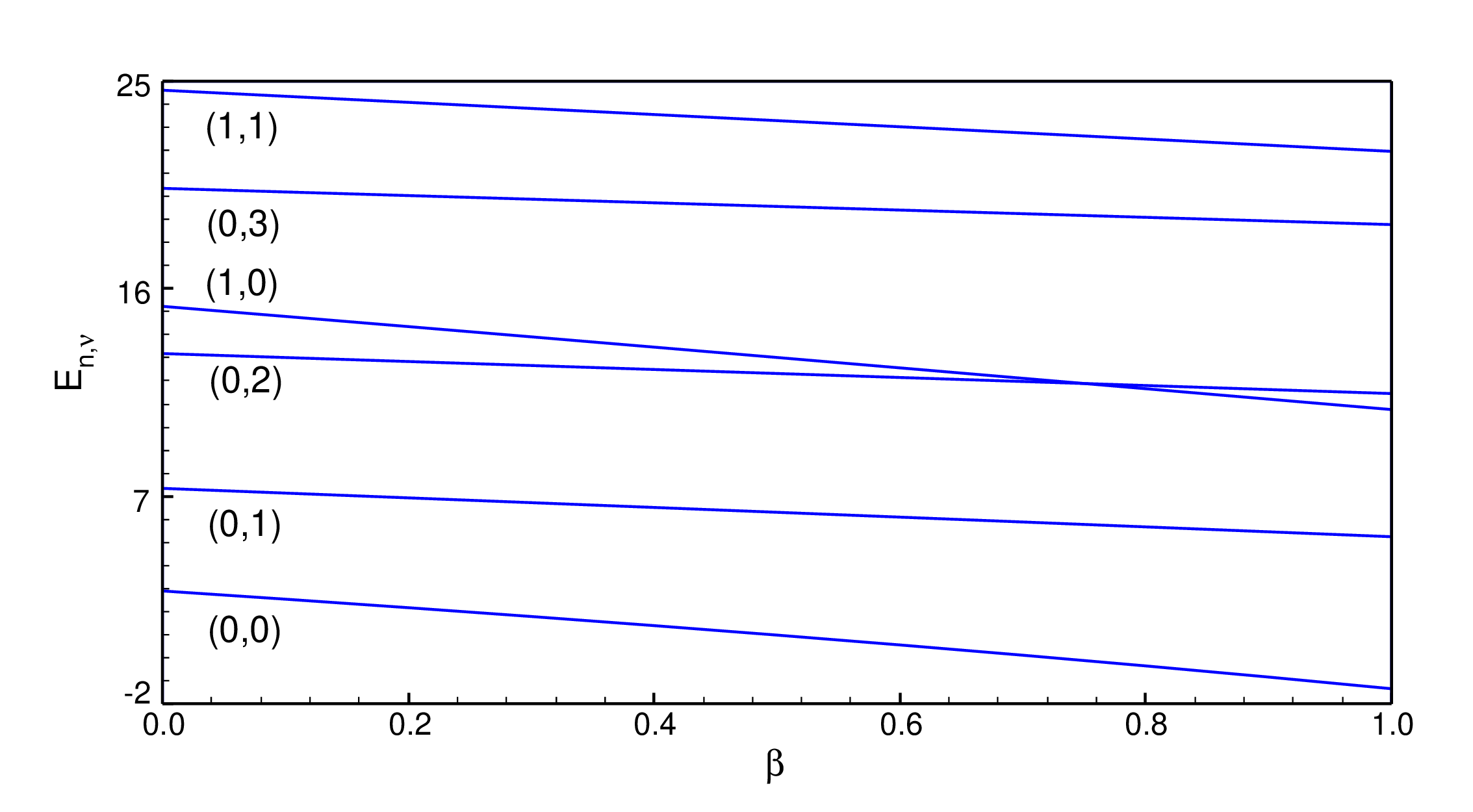}
\end{center}
\caption{Dimensionless eigenvalues $E_{n\nu}(\beta)$ for some values of $%
\beta$. Each line is labelled by the numbers $(n,\nu)$}
\label{Fig:E(beta)}
\end{figure}

Figure~\ref{Fig:E(beta)} shows the lowest eigenvalues $E_{n\nu }\left( \beta
\right) $ for $0\leq \beta \leq 1$. In this case, $n,\nu =0,1,\ldots $ are
good quantum numbers. We appreciate the crossing between the states $%
E_{02}\left( \beta \right) $ and $E_{10}\left( \beta \right) $ at $\beta
=0.75$ where the eigenvalue is three-fold degenerate. Since this particular
degeneracy does not follow from the symmetry arguments given in section~\ref
{sec:model} we consider it accidental and will come back to it in section~%
\ref{sec:f>0}.

It is worth noting that when $\beta =3/4$ we have the exact ground state
\begin{equation}
R_{00}(r)=\frac{(1-r)e^{(1-r)/2}}{\sqrt{3e-8}},\;E_{00}=-\frac{1}{8},
\label{eq:R_(00)_exact}
\end{equation}
that not only stisfies the boundary condition $R_{00}(1)=0$ but also $%
R(r\rightarrow \infty )=0$. Besides, our numerical results suggest that for
this particular value of $\beta $ $E_{n0}=E_{n-1\,2}$ for all $n=1,2,\ldots $%
. We will not try to prove this conjecture here because we are mainly
interested in the Stark effect discussed in section~\ref{sec:f>0}.

\section{Some results for $f\neq 0$}

\label{sec:f>0}

\begin{figure}[tbp]
\begin{center}
\includegraphics[width=9cm]{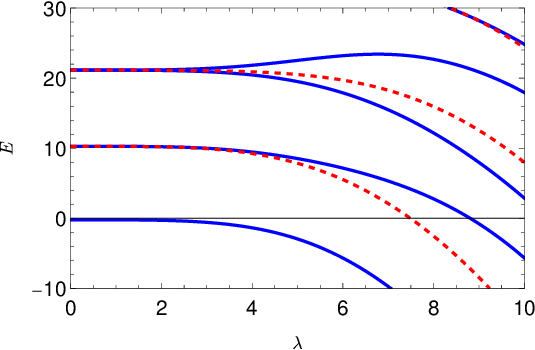}
\end{center}
\caption{Lowest eigenvalues for $r_0=3/4$ and some values of $\lambda$ }
\label{fig:Er034lamb}
\end{figure}

We have carried out calculations with the basis set (\ref{eq:f^eo})
including all the functions $i=0,1,\ldots ,12$ with the corresponding values
of $j$. Such basis sets provide sufficiently accurate eigenvalues for
present purposes. Figure \ref{fig:Er034lamb} shows the first eigenvalues for
$r_{0}=3/4$. We clearly appreciate the splitting of the two-fold degenerate
eigenvalue $E_{01}$ and the splitting of the anomalous three-fold degenerate
eigenvalue described in section~\ref{sec:f=0}.

\begin{figure}[tbp]
\begin{center}
\includegraphics[width=9cm]{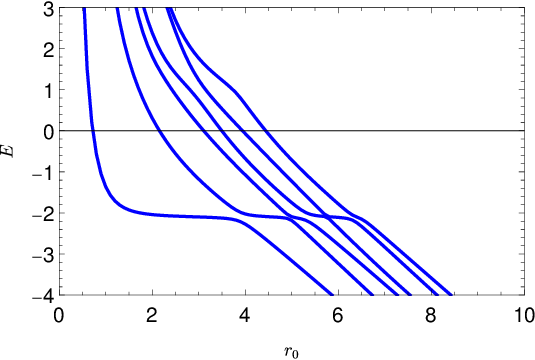}
\end{center}
\caption{Lowest eigenvalues for even states with $\lambda=1$ and some values
of $r_0$ }
\label{fig:Elamb1r0e}
\end{figure}

\begin{figure}[tbp]
\begin{center}
\includegraphics[width=9cm]{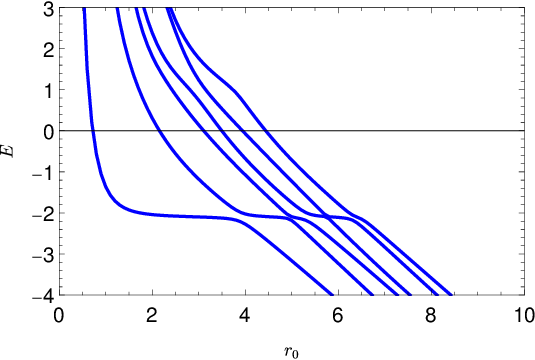}
\end{center}
\caption{Lowest eigenvalues for odd states with $\lambda=1$ and some values
of $r_0$ }
\label{fig:Elamb1r0o}
\end{figure}

Figures \ref{fig:Elamb1r0e} and \ref{fig:Elamb1r0o} show the first
eigenvalues for even and odd states, respectively, when $\lambda =1$. These
figures are also interesting because they show the existence of avoided
crossings that only take place between states of the same symmetry (see \cite
{F14} and references therein for a general discussion of avoided crossings).

\section{Conclusions}

\label{sec:conclusions}

In this paper we revisited a simple quantum-mechanical model discussed
earlier by other authors\cite{LGL23}. We restricted ourselves to the case in
which the nucleus is clamped at the origin of the circular box in order to
discuss several features of the solutions that have not been investigated
before. We analyzed the limits $r_{0}\rightarrow 0$ and $r_{0}\rightarrow
\infty $ by means of suitable dimensionless equations. The high symmetry of
the problem gives rise to an interesting behaviour of the eigenvalues, like
degeneracy in the absence of electric field. We have also discussed a case
of accidental symmetry that was not found in an earlier treatment of the
harmonic oscillator\cite{FGAF24}. Another interesting feature of this simple
model is the presence of avoided crossings that stem from the variation of
the box radius at constant nonzero values of the electric field.

\section*{Acknowledgements}

The research of P.A. was supported by Sistema Nacional de Investigadores
(M\'exico).

\begin{table}[htbp]
\caption{$r_0^2 E_{n,\nu}$ for some states $(n,\nu)$ with $\lambda=1$ and
two values of $r_0$}
\label{tab:r0^2E}{\tiny
\begin{tabular}{lD{*}{*}{0}D{*}{*}{0}D{*}{*}{0}D{*}{*}{0}D{*}{*}{0}D{*}{*}{0}}
\multicolumn{1}{c}{$r_{0}$} & \multicolumn{1}{c}{(0,0)}  &
\multicolumn{1}{c}{(0,1)} & \multicolumn{1}{c}{(0,2)} &
\multicolumn{1}{c}{(1,0)} & \multicolumn{1}{c}{(0,3)} &
\multicolumn{1}{c}{(1,1)} \\
0.01 & 2.856402592 & 7.320440988 & 13.17007975 & 15.19203641 &
20.33755843 &
24.58296830 \\
0.001 & 2.888078919 & 7.338931259 & 13.18558547 & 15.23127335 &
20.35166550 & 24.60660243 \\
\multicolumn{1}{c}{$E_{n,\nu}^{PB}$} & 2.891592981 &  7.340985321
&  13.18730821 &  15.23563117 &
20.35323290  & 24.60922816   \\

\end{tabular}
}
\end{table}

\end{document}